\documentstyle[12pt,aaspp4,psfig]{article}

\newcommand\eg{{\it e.g.} }
\newcommand\etal{et~al.}
\newcommand\lya{Ly$\alpha$}
\newcommand\CII{\hbox{C~$\rm II$}]}
\newcommand\CIV{\hbox{C~$\rm IV$}}
\newcommand\CIII{\hbox{C~$\rm III]$}}
\newcommand\Ha{H$\alpha$}
\newcommand\Hbeta{H$\beta$}
\newcommand\MgII{\hbox{Mg~$\rm II$}}
\newcommand\NeIII{\hbox{[Ne~$\rm III$}]}
\newcommand\NeIV{\hbox{[Ne~$\rm IV$}]}
\newcommand\NeV{\hbox{[Ne~$\rm V$}]}
\newcommand\OII{[\hbox{O~$\rm II$}]}
\newcommand\OIII{[\hbox{O~$\rm III$}]}
\font\aipsfont = cmsy9 scaled\magstep1

\newcommand\aips {{\aipsfont AIPS$\;$}}
\newcommand\ang{\AA}               
\newcommand\minpoint{$^{\prime}\mskip-4.7mu.\mskip0.8mu$}
\newcommand\secpoint{$^{\prime\prime}\mskip-7.6mu.\,$}

\begin{document}

\title{Discovery of an Ultra-Steep Spectrum, Highly Polarized Red Quasar at $z=1.462$\altaffilmark{1}}
\author{Carlos De Breuck\altaffilmark{2}, M. S. Brotherton, Hien D. Tran, Wil van Breugel}
\affil{Institute for Geophysics and Planetary Physics,
Lawrence Livermore National Laboratory, L$-$413, P.O. Box 808,
Livermore, CA 94550, U.S.A.}
\authoremail{debreuck@igpp.llnl.gov}
\author{Huub J. A. R\"ottgering}
\affil{Leiden Observatory, P.O. Box 9513, 2300 RA Leiden, The Netherlands}
\altaffiltext{1}{Based on observations obtained at the W.M. Keck Observatory, which is operated jointly by the University of California and the California Institute of Technology}
\altaffiltext{2}{Also at Leiden Observatory}

\begin{abstract}
We report the discovery of WN~J0717+4611, a highly polarized red quasar at $z=1.462$, selected from a sample of faint ($S_{1.4GHz} >$ 10 mJy) ultra steep spectrum ($\alpha_{1400}^{327} < -1.3$; $S \propto \nu^{\alpha}$) radio sources. VLA observations at 4.85 GHz show that the radio source is dominated by a 21 mJy core. Using the Keck telescopes, we obtained spectra with a combined wavelength range covering 4000~\ang\ to 1.6 \micron, and spectropolarimetric observations between 4000~\ang\ and 9000~\ang. We identify WN~J0717+4611 as a quasar based on broad emission of \MgII~$\lambda$2799 (FWHM $\simeq$ 11,000 km s$^{-1}$), a stellar $R_S$-band Lick image and the high luminosity of the object ($M_B=-24.2$). The optical linear polarization is wavelength-independent at 15\% in both the continuum and the emission lines, with the polarization angle orthogonal to the radio jet axis. We argue that the polarization mechanism is scattering of quasar light by surrounding small dust grains or electrons. WN~J0717+4611 appears highly reddened for a core-dominated quasar. 

\end{abstract}


\section{Introduction}

The tight correlation in the Hubble $K - z$ diagram of the parent galaxies of luminous radio sources suggests that high redshift radio galaxies (HzRGs) may be used to study the formation and evolution of massive ellipticals (\cite{lil89}, \cite{eal97}, \cite{wvb98}).
HzRGs have now been found out to redshifts of $\sim 4.5$, which is presumably close to the time they underwent their first burst of star formation. As a result of various vigorous search campaigns, $\sim 20$ HzRG are now known at $z > 3$.
Virtually all of these have been found by identifying sources with ultra-steep ($\alpha \lesssim -1$) radio spectra (\cite{cha96}; \cite{raw96}; \cite{rot97}; \cite{deb98a}). Although this ultra-steep spectrum (USS) search technique has now been proven the most effective way of finding HzRGs, no uniform large-area search could be done because the radio surveys used to construct the samples were either not sensitive enough, or had limited sky coverage. During the past years, several new, deep all-sky radio surveys have become available: the Faint Images of the Radio Sky at Twenty Centimeters at 1.4 GHz (FIRST, \cite{bec95}), the Westerbork Northern Sky Survey at 325 MHz (WENSS, \cite{ren97}), and the NRAO VLA Sky Survey at 1.4 GHz (NVSS, \cite{con98}). These allow for the first time a major survey of faint radio sources with extreme spectral index properties.

We have used these new surveys to define a sample of USS sources, for further optical identification and spectroscopic studies (see \cite{deb98a} for a summary). Our sample includes sources with radio flux densities $\sim10 \times$ fainter than previous large USS searches (\cite{cha96}; \cite{rot97}).
Subsequent optical imaging and spectroscopy of our sample indicates that 2/3 of the USS sources with $R < 24$ are radio galaxies with redshifts $z > 2$. We will report in detail on these results in forthcoming papers.

In this paper, we present the discovery and follow-up observations of one of our USS sources, WN~J0717+4611 at $z = 1.462$. This source was selected for a special study because of its strong core-dominated FR II morphology (\cite{far74}) and a bright stellar $R_S = 21.6$ identification. Such a combination of properties is unique in our USS sample, and its bright compact optical identification is reminiscent of 6C 1908+722, a $R = 21$ radio galaxy at $z = 3.515$ with a broad absorption line (BAL) system (\cite{dey97}).

Our Keck optical spectropolarimetry observations do not reveal such a BAL system, but, in combination with the IR-spectroscopy that was originally obtained to study the age of the stellar population by means of the strength of the 4000~\ang-break, it shows evidence for a highly polarized red quasar. We argue that the polarization is caused by scattering and not by synchrotron radiation, as in most highly polarized quasars (\cite{coh97}). The spectral properties of the polarized light indicates that the red color of WN~J0717+4611 is strong evidence for dust obscuration, which has been one of the prime explanations for the red quasars (\eg \cite{web95}), although other mechanisms, like a host galaxy or synchrotron contribution, have also been proposed (\eg \cite{ben98}).

We describe the radio and optical imaging, IR spectroscopy, and optical spectropolarimetry in \S\ 2, and present the results in \S\ 3. We discuss the classification, the origin of the high polarization, and compare WN~J0717+4611 with possibly related objects in \S\ 4, and present our conclusions in \S\ 5. Throughout the paper, we use $H_0 = 50$ km s$^{-1}$ Mpc$^{-1}$, $q_0 = 0$ and $\Lambda = 0$. At $z = 1.462$, changing these values to $H_0 = 65$ km s$^{-1}$ Mpc$^{-1}$, $q_0 = 0.1$, would reduce size and distance by 30\%, which would not affect our conclusions.

\section{Observations and data reduction}

\subsection{Radio Observations}

Table 1 summarizes the radio flux density measurements for WN~J0717+4611. Our USS sample includes all sources with $\alpha_{1400}^{327} < -1.3$, determined from the WENSS 327 MHz and NVSS 1.4 GHz radio surveys. Both these surveys have comparable resolutions of 54\arcsec\ and 45\arcsec, respectively, minimizing missing flux problems in the spectral index determination.
WN~J0717+4611 has a spectral index of $\alpha_{1400}^{327} = -1.31$ which satisfies the above selection criterion. For a full discussion of the sample and selection techniques, see \cite{deb98b}.

To determine the radio morphology and accurate astrometry for optical identification, we obtained snapshot VLA observations of our USS sample on 1996 October 28 with the A-array.
As part of this program, WN~J0717+4611 was observed for 5 minutes at a frequency of 4.86 GHz at 0\farcs4 resolution.
We used standard data reduction techniques in the NRAO \aips package, including phase-only self-calibration.
3C 286 was the primary flux calibrator; the difference between the predicted and observed flux of the other observed flux calibrator (3C 48) indicates that our flux calibration is accurate to within 2\%.
The final map (Fig.\ \ref{debreuck.fig1}) has an rms noise level of 86 $\mu$Jy, and shows an FR II morphology with an unusually bright core, located at RA(J2000) = 7$^h$17$^m$58\fs48, DEC(J2000) = +46\arcdeg11\arcmin38\farcs9. The southern lobe is edge-brightened and about a factor of 2 brighter than its northern, more diffuse, counterpart. To check the structure of these lobes and to improve the sensitivity to detect low-brightness structures, we also mapped the source using only the core for self-calibration, and limiting the UV-coverage to half the maximum range. All maps show the same morphology seen in Figure \ref{debreuck.fig1}.

\placefigure{debreuck.fig1}

\subsection{Lick Identification and Spectroscopy}

$R_S-$band (\cite{djo85}) imaging and long-slit spectroscopic observations were made with the KAST double-beam imaging spectrograph (\cite{mil94}) on the Lick 3m telescope on 1996 November 11, under photometric conditions and a seeing of 1\farcs3.
We determined the astrometric solution of the image with the \aips task XTRAN, using coordinates of 8 stars detected on the POSS plates. The resulting astrometric accuracy is $\sim$ 1\arcsec.
A stellar $R_S = 21.6 \pm 0.3$ object was detected at the location of the radio core (Fig.\ \ref{debreuck.fig1}).
A 20-minute spectrum was taken with a 2\arcsec~ slit oriented east-west, centered on the object, and using a dichroic that splits the light at 5500~\ang.
We used a 452/3306 grism and a 300/7500 grating in the blue and red arm resulting in a dispersion of 2.54~\ang/pix (resolution $\sim$ 6~\ang\ FWHM) and 4.6 ~\ang/pix (resolution $\sim$ 8~\ang\ FWHM), respectively, covering a continuous wavelength range from 3500~\ang\ to 10,000~\ang.
The spectrum (not shown) shows a weak continuum in the red and no continuum and a single narrow line in the blue at 3816~\ang\ . We initially assumed this line to be \lya~ from a radio galaxy at $z=2.138$, because of the absence of \Hbeta\ and \OIII~$\lambda\lambda$4959, 5007 if the identification would be \OII~$\lambda$3727. The Keck optical spectrum (\S 2.4) showed that the correct identification of this line is \CIV~$\lambda$1549.

\subsection{Keck Infrared Spectroscopy}

A low-resolution IR spectrum was obtained on 1996 November 29 using the Near Infra-Red Camera (NIRC; \cite{mat94}) at the 10m Keck I telescope. The gr150 grism was used in combination with the $JH$-filter. The dispersion was $\sim$ 50~\ang/pix, and the slit width was 1\farcs275, resulting in a resolution of 190~\ang\ (FWHM). Conditions during the observations were photometric, but the seeing was variable. To achieve good sky-subtraction, we used ten 5-minute exposures, while shifting the object along the slit. The first six exposures had poor seeing, and were discarded. The 5 minute integrations were slightly too long under these variable atmospheric conditions, resulting in saturated sky lines between 1.45 and 1.6\micron.

To remove dark counts and sky emission, we subtracted the average of the two bracketing exposures from each observation. We divided the resulting frames by the normalized domeflat, and removed the residual sky emission by subtracting a low-order fit along the spatial direction.
We then shifted the four remaining good frames and combined them, while masking out the bad pixels determined from the domeflats.

For wavelength calibration, we initially used the dispersion formula listed for the gr150 grism in the NIRC manual. This was further refined by assuming the redshift determined from spectropolarimetric observations (\S\ 2.4), and shifting the \Hbeta~ line by 120 \ang\ (60\% of resolution) to match this redshift.

We observed HD~56605 (an A3 star to minimize intrinsic spectral features), immediately after WN~J0717+4611. These observations were reduced in the same manner as our primary target. After extraction of the 1-dimensional spectra, WN~J0717+4611 was divided by HD~56605 to correct for atmospheric absorption.
HD~56605 was also used for absolute flux calibration, under the assumptions that it has a main sequence color $V-J$ = +0.15, a blackbody temperature $T_{eff}$ = 9000K (\cite{jon66}), and $V = 8.0$ (SIMBAD database).

The flux-calibrated spectrum matches up well with the optical spectrum described below, confirming the reliability of the flux calibration and the photometric conditions during both runs.

\subsection{Keck Spectropolarimetry}

The spectropolarimetric observations were made on 1996 December 9 at the 10m Keck II telescope using the Low Resolution Imaging Spectrometer (LRIS, \cite{oke95}) in conjunction with the polarimeter (\cite{coh97}). The observations were split into four 15-minute exposures, each at a different orientation of the waveplate at angles of 0\arcdeg, 45\arcdeg, 22\fdg5, and 67\fdg5 (see \cite{goo95} for details).
The 300 l mm$^{-1}$ grating blazed at 5000\ang\ combined with the 1\arcsec~ wide slit resulted in a resolution of $\sim$ 10\ang\ (FWHM) .
The dispersion was 2.5~\ang/pix, yielding a spectral coverage from 4015~\ang\ to 9005~\ang\ .
No order blocking filter was used, since second order contamination from the blue is negligible, as the spectrum is very red. Atmospheric conditions were photometric, but the seeing was poor and variable (up to 1\farcs5), making the spectrophotometry with the 1\arcsec~ slit unreliable.

The observations were reduced using the NOAO IRAF package and SuperMongo routines developed by \cite{dey96}, based on the method described by \cite{mil88}.
The flux calibration was performed using the standard star Feige~34 (\cite{mas88}). The polarization measured from this unpolarized star was $<0.15$ \% at all wavelengths, validating our polarimetric calibration. 
Our polarimetric results are presented in Figure \ref{debreuck.fig3}. The faint continuum necessitated binning of the Stokes $Q$ and $U$ data into $\sim$1000~\ang\ bins to improve S/N. To check the validity of this result, we used different bin sizes, which were all consistent with the results shown in Figure \ref{debreuck.fig3}.

\placefigure{debreuck.fig3}

We corrected the spectrum for telluric A and B absorption (due to O$_2$) using an atmospheric transmission curve made by dividing the flux calibrated spectrum of Feige~34 by a model spectrum interpolated over the A and B bands. The atmospheric conditions during the target and standard star observations were similar, but there was a significant difference in airmass between Feige~34 (1.1) and WN~J0717+4611 ($\sim$1.5).
The B band coincides with the narrow-line part of the \MgII~$\lambda$2799 region. We estimate that the residual error after B band correction is $\sim$ 20\% and this could significantly affect our measurement of the width and strength of the narrow component. The broad component of \MgII~$\lambda$2799 is not affected by the B band, as it is three times broader than the B band.

\placefigure{debreuck.fig2}

\section{Results}

The radio-spectrum (Fig.\ \ref{debreuck.fig2}) flattens with increasing frequency, with the spectral index changing from $\alpha = -1.59$ between 327 MHz and 408 MHz to $\alpha = -0.84$ between 1.4 GHz and 4.85 GHz. This suggests an increasing contribution with frequency of a flat spectrum component, probably the core.

The combined total intensity spectrum from the Lick and Keck observations (Fig.\ \ref{debreuck.fig4}) shows 11 emission lines and a red continuum with $\alpha_{opt}=-1.6 \quad (F_{\lambda} \propto \lambda^{\alpha_{opt}})$. We used the SPECFIT package inside IRAF to fit the different components of the spectrum, and present the measurements in Table 2.
The average redshift of WN~J0717+4611 is $z=1.462\pm 0.001$. We do not detect any significant velocity shifts among the narrow emission lines.
The \MgII~$\lambda$2799 line appears to have a $\sim$350 km s$^{-1}$ redshift relative to the other lines, but this can be accounted for by using a higher multiplication factor for the atmospheric B band correction, as suggested by the residuals at the A band and the higher airmass of the object.
All emission lines except \MgII~$\lambda$2799 show only a narrow (FWHM $\sim$ 1200 km s$^{-1}$) component. In \MgII~$\lambda$2799, we detect 3 components: a narrow (FWHM = 1200 km s$^{-1}$) component, a broad (FWHM = 11,000 km s$^{-1}$) component, and an intervening foreground absorption system blueshifted by 8500 km s$^{-1}$ at $z=1.395$ (Fig.\ \ref{debreuck.fig3}).

A comparison of the smoothed version of the Lick spectrum and the Keck total intensity optical spectrum shows a close agreement in both continuum slope and level, suggesting that (i) conditions during both nights were photometric, (ii) the continuum emission from the object originates from an area that is $\lesssim$ 1\arcsec, and (iii) the object is not highly variable in the optical on timescales of $\sim$ 1 month. The overlay also explains why we did not previously detect the \MgII~$\lambda$2799 in the Lick spectrum: the stronger atmospheric B band absorption at Lick Observatory reduced the amplitude of this line to $\sim 1 \sigma$.

The low resolution and large errors in the wavelength calibration of the IR spectrum make it impossible to determine the velocity shifts of the lines. The \NeIII~$\lambda$3869 and \Ha~ lines are both detected at the edges of the spectrum, but the fluxes and widths of both lines are highly uncertain. The only two lines in a reliable part of the IR spectrum are \Hbeta\ and the \OIII~$\lambda\lambda$4957, 5007 doublet. However, at $z = 1.462$, the separation between \Hbeta\ and \OIII~$\lambda$4959 is 240~\ang\ --- barely resolved with our 190~\ang\ resolution. We therefore used the IR spectrum only to confirm the redshift and constrain the reddening of the continuum.

We detect a percentage polarization of 15$\pm$3\% and a polarization angle of 110\arcdeg$\pm$10\arcdeg, both constant with wavelength within the uncertainties. The polarization angle is orthogonal to line connecting the hotspots with the radio core. The percentage polarization and polarization angle are similar in the emission lines and the continuum.

\section{Discussion}
Since the object was not at the initially assumed redshift, the IR spectroscopy and optical spectropolarimetry were not appropriate to detect a 4000~\ang-break nor a \CIV~ BAL system. However, this data showed that this object is a red quasar with a highly polarized optical continuum, which we will argue in the context of Unified Models, presents evidence for scattering by dust or electrons.
 
\subsection{Classification}

We classify WN~J0717+4611 as a quasar because:

\begin{enumerate}
\item{We detect a broad component (FWHM $\simeq$ 11,000 km s$^{-1}$) underlying the narrow \MgII~$\lambda$2799 emission line (Fig.\ \ref{debreuck.fig3}).}

\item{The Lick $R_S-$band image shows a stellar ($\lesssim$ 1\farcs3) $R_S \simeq$ 21.6 object coinciding with the radio core.}

\item{The absolute magnitude $M_B = -24.2$ is brighter than the $M_B = -23$ cutoff used by \cite{ver96} to classify stellar objects as quasars. The $M_B$ has been calculated assuming a constant power law with index $\alpha_{opt} = -1.6$ without line-contamination correction, but has not yet been corrected for Galactic or intrinsic reddening, and should therefore be interpreted as a lower limit on the luminosity.}

\item{Of the 28 objects in our USS sample for which we have already obtained images and spectra, only three are quasars on the basis of broad lines (including WN~J0717+4611). All three have an unresolved, $R \sim 21$ core identification, while the other 25 are radio galaxies which show an extended, $R \gtrsim 23$ optical counterpart.}
\end{enumerate}

As mentioned in \S\ 3, the object did not show any optical variability within a timescale of $\sim$ 1 month. WN~J0717+4611 is not an optical violently variable (OVV; \cite{jan93}), as most highly polarized quasars are (\cite{jan93}). Further reasons to exclude WN~J0717+4611 from being an OVV are the extremely steep radio spectrum (\cite{gea94}), and the rejection of the synchrotron mechanism as the explanation of the high polarization (\S\ 4.2).

The radio emission from WN~J0717+4611 is strongly core-dominated, and we derive a rest-frame 12 GHz radio-core to lobe ratio $R \simeq 0.6$ (\cite{orr82}; \cite{wil95}). Baker \& Hunstead (1995; hereafter BH95) have classified radio-loud quasars based this core dominance parameter, and constructed composite spectra for different values of $R$. One of the main results was that both the strength of the Fe line complex near \MgII~$\lambda$2799 and the amount of reddening increase with decreasing core dominance (smaller $R$).
This can be understood in the context of the AGN unification model (\eg \cite{ant93a}), in which the fainter radio cores are thought to be central AGNs that are only modestly Doppler boosted since their jets are at relatively large angles of the line of sight. As a result, such quasars are thought to experience a higher reddening by the dusty torus surrounding the AGN (\cite{bak97}).
On the basis of the core-dominance and the \cite{bak95} classification, we would statistically expect relatively strong Fe-lines, and low reddening. This is contrary to what is observed, and in fact, a much better similarity (Fig. \ref{debreuck.fig4}) is found when comparing our combined optical--IR spectrum with the average spectrum of Compact Steep Spectrum (CSS) radio sources in \cite{bak95}.

In the top panel of Figure \ref{debreuck.fig4}, we show the observed spectra, corrected for Galactic reddening with an $A_V = 0.25$ [value as listed in the NED database, determined from \cite{bur82} and using the \cite{car89} extinction curve].
We subsequently dereddened this spectrum with an $A_V = 1.2$ to match to CSS composite, assuming an SMC-type extinction curve (\cite{pre84}) at the redshift of the quasar. We used the SMC type law rather than the Galactic, because no 2200~\ang\ dust feature was obvious (\cite{bon96}). The resulting spectrum, presented in the bottom panel of Figure \ref{debreuck.fig4} shows a close resemblance with the \cite{bak95} composite CSS spectrum. However, the physical size of the radio lobes of WN~J0717+4611 is 92 kpc, which is significantly larger than the commonly accepted maximum size of 20 kpc for CSS sources (\eg \cite{san95}). Why then do we see this resemblance?
\cite{bak95} suggest a significant reddening might explain the extremely red continuum in CSS sources compared to the more core-dominated quasars. A straightforward interpretation of the resemblance of WN~J0717+4611 to the CSS composite would therefore be high reddening by dust in both. 
Similarly, high amounts of reddening have been seen in other radio-loud quasars (\eg IRAS 13349+2438, \cite{wil92}; PKS 1610$-$771, \cite{cou97}; 5C7.195, \cite{wil98a}), and have been proposed to hide a significant population of quasars (\cite{web95}).

\placefigure{debreuck.fig4}

\subsection{The Origin of the Polarization}

We consider the four most common mechanisms that can produce the observed polarization.

\begin{enumerate}
\item{{\bf Optical synchrotron emission:}
A significant contribution to the continuum polarization by optical synchrotron emission can be excluded because the degree of polarization and polarization angle of the narrow emission lines (which cannot be caused by synchrotron radiation) are similar to those in the continuum, implying a common non-synchrotron origin.}

\item{{\bf Transmission through aligned dust grains:}
Another possible polarization mechanism is transmission of nuclear light through aligned dust grains. However, the high 15\% polarization, and the flat shape of the wavelength dependence cannot be fit with a Serkowski law (\cite{ser75}). Furthermore, in our Galaxy, the amount of interstellar polarization is related to the amount of interstellar reddening by $P_{max} < 9 \times E(\bv)\%$, which would imply an $A_V > 5$ in WN~J0717+4611, much higher than observed.

For similar reasons, the polarization cannot be due to a foreground screen in our own Galaxy: the observed $E(\bv) = 0.1$ and at the Galactic latitude of WN~J0717+4611 ($b\sim24$\arcdeg) no polarization at the 15\% level has been detected (\cite{ser75}).}

\item{{\bf Scattered AGN light by dust or electrons:}
The most likely mechanism is scattering by dust or electrons. The mean polarization angle of 110\arcdeg\ is perpendicular to the line connecting the 2 hotspots through the core. This is just what is expected in the AGN unification model, where we would be observing continuum emission from the AGN being scattered within the cone containing the radio jets.
Both the high level and the wavelength-independence of the polarization can be explained with scattering by electrons or small ($2\pi a / \lambda \ll 1$) dust grains (\eg \cite{wil92}).}
\end{enumerate}

\subsection{Comparison with Other Quasars}

Highly Polarized Quasars (HPQs) are defined as quasars with an optical polarization $P >$ 3\%, independent of the polarization mechanism (\eg \cite{jan93}). Because these objects are rare, and the surveys used to find them were biased, no general properties of HPQs have been determined. We therefore briefly compare WN~J0717+4611 with other HPQs in the literature that have similar characteristics.

Another radio-loud HPQ that is not an OVV is OI~287 ($z = 0.445$, \cite{goo88}; \cite{ant93b}). This object has a continuum polarization of 8.2\%, independent of wavelength. However, its polarization properties are different from WN~J0717+4611: OI~287 has narrow emission lines that are consistent with being unpolarized, but the broad \Hbeta~ does show polarization, while in WN~J0717+4611 the emission lines are polarized at the same level as the continuum. Furthermore, the polarization angle of the continuum is parallel to the radio structure, while it is orthogonal in WN~J0717+4611. Hence it has been suggested (\cite{goo88}) that OI~287 has a thin obscuring disk between the NLR and BLR. In the case of WN~J0717+4611, the obscuring torus seems to be large and encompass also the NLR.

An object that is very similar to WN~J0717+4611 in its optical properties is 3CR~68.1 ($z = 1.228$; \cite{bro98}). 3CR~68.1 is also an HPQ that is neither an OVV nor BALQSO. The optical polarization is also caused by scattering, as in WN~J0717+4611, and it is also an extremely red quasar. In the context of the unified schemes, it is argued that 3CR~68.1 is seen under a dusty, highly inclined view. The radio properties are very different from WN~J0717+4611 however, as 3CR~68.1 is an extremely lobe-dominated source ($R \sim 4 \times 10^{-4}$) with the weakness of its core possibly due to free-free absorption. We interpret WN~J0717+4611 as a less inclined object than 3CR~68.1, but with a similar amount of dust obscuration and scattered light.

The object that is perhaps most similar to WN~J0717+4611 is 5C~7.195 ($z$ = 2.034; \cite{wil98a}; \cite{wil98b}). Its radio structure also shows a core-dominated FR-II morphology and has a steep radio spectrum. Similarly, the optical object is unresolved, and its spectrum can be fitted by a quasar spectrum reddened by $A_V = 1.1$. The quasar nature is also supported by the detection of broad \Ha. As Willot \etal\ (1997) remark, a significant number of these red quasars could easily have been classified as radio galaxies, as the underlying broad lines require good seeing and high S/N measurements. WN~J0717+4611 is a perfect example: without the high-quality Keck spectrum, we would have classified this object as a radio galaxy.

\section{Conclusions}
Our main observational results for WN~J0717+4611 are:

1) The radio source has a core-dominated FR II morphology, with a total size of $\sim 90$~kpc and an ultra-steep spectrum with an integrated spectral index $\alpha_{1400}^{327} = -1.31$. The radio spectrum flattens with increasing frequency, presumably because of an increasing contribution from a flatter spectrum core.

2) The optical object is a quasar because it is stellar, has broad \MgII~$\lambda$2799 emission with (FWHM $\simeq$ 11,000 km s$^{-1}$), and is very luminous with $M_B = -24.2$.

3) The object appears exceptionally reddened for its core-dominance of $R \simeq 0.6$, compared to the statistical results of \cite{bak95}. When dereddened with an $A_V = 1.2$, it matches well with the composite CSS quasar spectrum, which is already more reddened than the core-dominated quasar composite.

4) We detect a wavelength-independent 15\% optical polarization, which is similar in both continuum and emission lines. The polarization angle is orthogonal to the radio axis, which, in the context of the unified models of quasars and radio galaxies, is consistent with scattering due to small dust grains or electrons located in the cone containing the radio jet. 

We have compared WN~J0717+4611 with other HPQs and red quasars in the literature, and found no object that shares all observational characteristics. 5C1.195 (\cite{wil98a}) seems to be a very similar object, and it was discovered in a similar way, indicating more of these objects may be found from radio-selected samples. 
The presence of a significant population of (highly) reddened quasars at cosmological redshifts is currently subject to considerable debate (\cite{web95}; \cite{ben98}). While our observations of a single object do not bear any statistical significance on the size of this population, they do show that highly reddened, dusty quasars do indeed exist.

\acknowledgments

We thank Arjun Dey, Adam Stanford and Dan Stern for providing the Keck
data and their advise on the data reduction, Andrea 'Grapes' Cimatti
and Chris O'Dea for useful discussions, and Jo Baker for providing the
composite Molonglo quasar spectra. The W.M. Keck Observatory is a
scientific partnership between the University of California and the
California Institute of Technology, made possible by a generous gift
of the W.M. Keck foundation. The VLA is a facility of the National
Radio Astronomy Observatory, which is operated by Associated
Universities Inc. under cooperative agreement with the National
Science Foundation.  This research has made use of the NASA/IPAC
Extragalactic Database (NED) which is operated by the Jet Propulsion
Laboratory, California Institute of Technology, under contract with
the National Aeronautics and Space Administration. Work performed at
the Lawrence Livermore National Laboratory is supported by the DOE
under contract W7405-ENG-48.

\newpage

\begin{deluxetable}{lrrrr}
\tablewidth{0pc}
\tablecaption{Radio Data}
\tablehead{
\colhead{Catalog}      &\colhead{Frequency}      & \colhead{Flux\tablenotemark{a}} &
\colhead{Resolution} & \colhead{Reference} \\
\colhead{} & \colhead{(MHz)}          & \colhead{(mJy)} &
\colhead{}      & \colhead{} }

\startdata
WENSS & 327 $\quad$ & 724 $\pm$ 15$\;$ & 54\arcsec $\times$ 78\arcsec & \cite{ren97} \nl
TEXAS & 365 $\quad$ & 553 $\pm$ 82\tablenotemark{b}$\;$ & 20\arcsec $\times$ 28\arcsec & \cite{dou96} \nl
B3 & 408 $\quad$ & 510 $\pm$ 60$\;$ & 2\minpoint6 $\times$ 4\minpoint8 & \cite{fic85} \nl
NVSS & 1400 $\quad$ & 105 $\pm$ 10 & 45\arcsec $\times$ 45\arcsec & \cite{con98} \nl
GB6 & 4850 $\quad$ & 24 $\pm$ 12$\;$ & 3\minpoint7 $\times$ 3\minpoint3 & \cite{gre91} \nl
&&&&\nl
VLA (total) & 4860 $\quad$ & 36.9 $\pm$ 0.3 & 0\secpoint43 $\times$ 0\secpoint39 & This paper \nl
VLA (lobes) & 4860 $\quad$ & 15.7 $\pm$ 0.3 & 0\secpoint43 $\times$ 0\secpoint39 & This paper \nl
VLA (core) & 4860 $\quad$ & 21.2 $\pm$ 0.3 & 0\secpoint43 $\times$ 0\secpoint39 & This paper \nl
\enddata
\tablenotetext{a}{Errors quoted are 3$\sigma$}
\tablenotetext{b}{This source has an environment flag ``C'', indicating that the fringe amplitudes were affected by surrounding sources (see \cite{dou96} for details)}
\end{deluxetable}

\begin{deluxetable}{llrrrrrr}
\tablewidth{0pc}
\tablecaption{Line Measurements}
\footnotesize
\tablehead{
\colhead{Instrument \tablenotemark{a}} &
\colhead{Line}      & \colhead{$\lambda_{rest}$} &
\colhead{{$\lambda_{obs}$}} & \colhead{z$_{em}$} &
\colhead{F$_{int}$\tablenotemark{b}} & \colhead{FWHM\tablenotemark{c}} &
\colhead{W$_{\lambda}^{rest}$} \\
\colhead{} &
\colhead{}          & \colhead{(\ang)} &
\colhead{(\ang)}          & \colhead{} &
\colhead{(10$^{-16}$ erg s$^{-1}$ cm$^{-2}$)} & \colhead{(km s$^{-1}$)} &
\colhead{(\ang)}}

\startdata
KAST & \CIV & 1549.1 & 3816 $\pm$ 2$\;\>$ & 1.463$\;$ & 30.2 $\pm$ 4.0$\qquad$ & 1200 $\pm$ 400 & 28 $\pm$ 12$\;$ \nl
LRIS & \CIII & 1908.7 & 4700 $\pm$ 2$\;\>$ & 1.462$\;$ & 0.9 $\pm$ 0.1$\qquad$ & 1200 $\pm$ 300 &  6 $\pm$ 1$\;\;\>$ \nl
LRIS & \CII & 2326.8 & 5726 $\pm$ 2$\;\>$ & 1.461$\;$ & 0.4 $\pm$ 0.1$\qquad$ & 1200 $\pm$ 400 & 3.2 $\pm$ 0.4 \nl
LRIS & \NeIV & 2425.0 & 5968 $\pm$ 2$\;\>$ & 1.461$\;$ & 0.6 $\pm$ 0.1$\qquad$ & 1000 $\pm$ 300 & 2.8 $\pm$ 0.4 \nl
LRIS & \MgII (narrow) & 2799.1 & 6900 $\pm$ 2$\;\>$ & 1.465$\;$ & 2.0 $\pm$ 0.4$\qquad$ & 1300 $\pm$ 300 & 13 $\pm$ 1$\;\;\>$ \nl
LRIS & \MgII (broad) & 2799.1 & 6900 $\pm$ 1$\;\>$ & 1.465$\;$ & 4.6 $\pm$ 0.4$\qquad$ & 11000 $\pm$ 500 & 13 $\pm$ 1$\;\;\>$ \nl
LRIS & \MgII (abs.) & 2799.1 & 6703 $\pm$ 1$\;\>$ & 1.395\tablenotemark{d}$\;$ & \nodata$\;\>$ $\qquad$ & 2000 $\pm$ 300 & $-$7 $\pm$ 1$\;\;\>$ \nl
LRIS & \NeV & 3345.9 & 8242 $\pm$ 3$\;\>$ & 1.463$\;$ & 0.5 $\pm$ 0.3$\qquad$ & 800 $\pm$ 400 & 1.2 $\pm$ 0.4 \nl
LRIS & \NeV & 3426.0 & 8437 $\pm$ 2$\;\>$ & 1.463$\;$ & 0.7 $\pm$ 0.3$\qquad$ & 600 $\pm$ 300 & 1.2 $\pm$ 0.4 \nl
NIRC & \NeIII & 3868.8 & 9500 $\pm$ 50 & \nodata$\;$ & \nodata$\;\>$ $\qquad$ & \nodata$\;\;\;\,$ & \nodata$\;\;\>$ \nl
NIRC\tablenotemark{e} & \Hbeta & 4861.3 & 11940 $\pm$ 30 & \nodata$\;$ & 11 $\pm$ 5$\;\>$ $\qquad$ & \nodata$\;\;\;\,$ & 45 $\pm$ 10$\;$ \nl
NIRC\tablenotemark{e} & \OIII & 4959.5 & 12230 $\pm$ 30 & \nodata$\;$ & 18 $\pm$ 5$\;\>$ $\qquad$ & \nodata$\;\;\;\,$ & 70 $\pm$ 15$\;$ \nl
NIRC\tablenotemark{e} & \OIII & 5006.8 & 12350 $\pm$ 30 & \nodata$\;$ & 53 $\pm$ 5$\;\>$ $\qquad$ & \nodata$\;\;\;\,$ & 220 $\pm$ 15$\;$ \nl
NIRC & \Ha & 6562.8 & 16265 $\pm$ 50 & \nodata$\;$ & \nodata$\;\>$ $\qquad$ & \nodata$\;\;\;\,$ & \nodata$\;\;\>$ \nl
\enddata
\tablenotetext{a}{KAST = KAST on Lick 3m, 2\arcsec slit; LRIS = LRIS on Keck II, 1\arcsec slit; NIRC = NIRC JH filter + gr150 grism on Keck I, 1\secpoint275 slit}
\tablenotetext{b}{Corrected for Galactic reddening using the \cite{car89} extinction curve with A$_V$=0.24}
\tablenotetext{c}{Deconvolved with the instrumental resolution (5.5\ang\ for KAST; 10\ang\ for LRIS)}
\tablenotetext{d}{Absorption line system with a blue velocity shift of 8500 km s$^{-1}$}
\tablenotetext{e}{FWHM unresolved with the 4600 km s$^{-1}$ resolution}
\end{deluxetable}

\newpage
\psfig{file=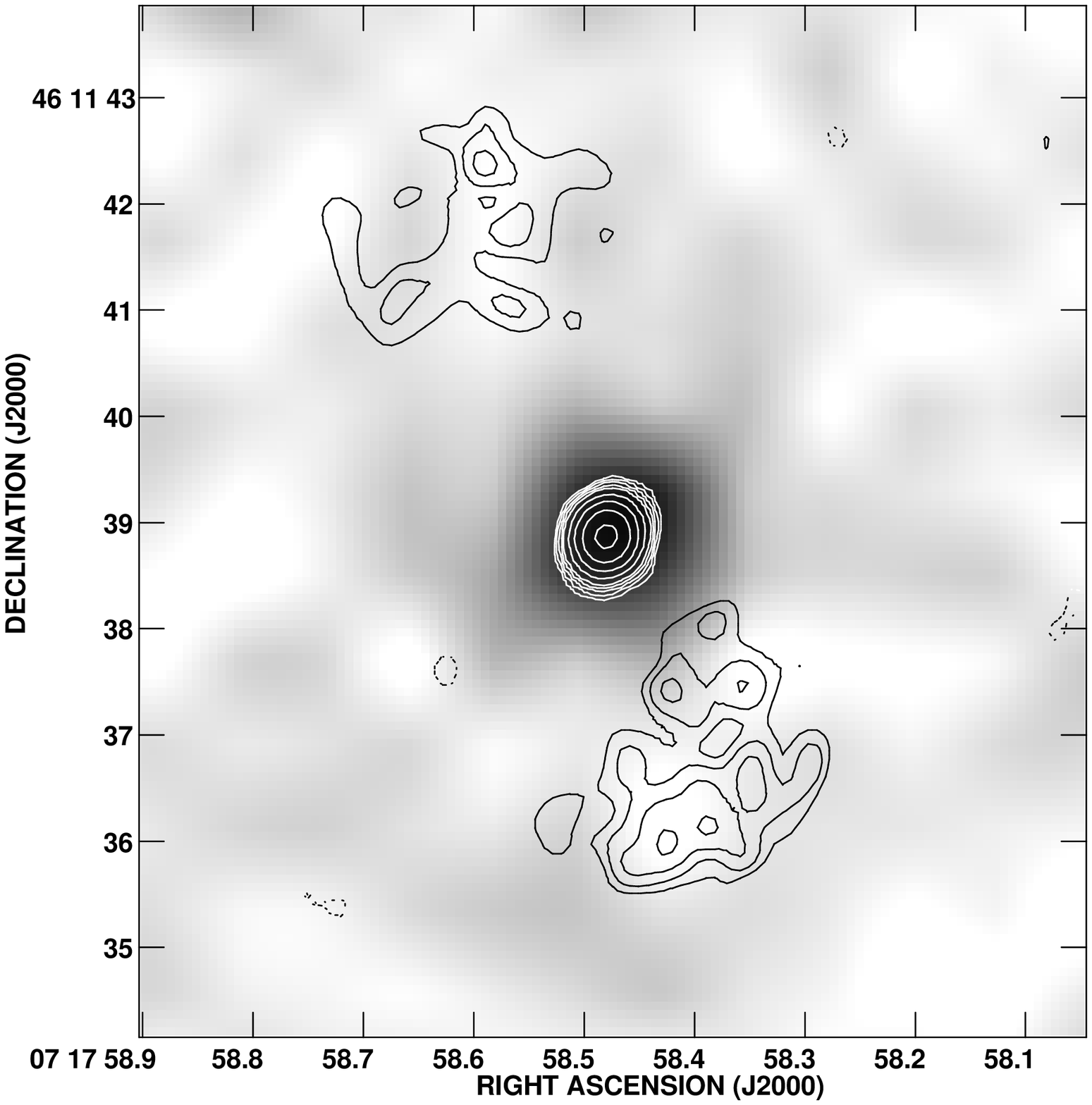,width=16cm}
\figcaption[debreuck.fig1.ps]{Greyscales: R$_S-$band Lick image; Contours: VLA 4.86 GHz image. Contour levels are $-$0.26, 0.26, 0.37, 0.52, 0.73, 1.0, 1.5, 2., 2.9, 4.1, 5.8, 8.3, 11.7 and 16.5 mJy beam$^{-1}$. \label{debreuck.fig1}}

\psfig{file=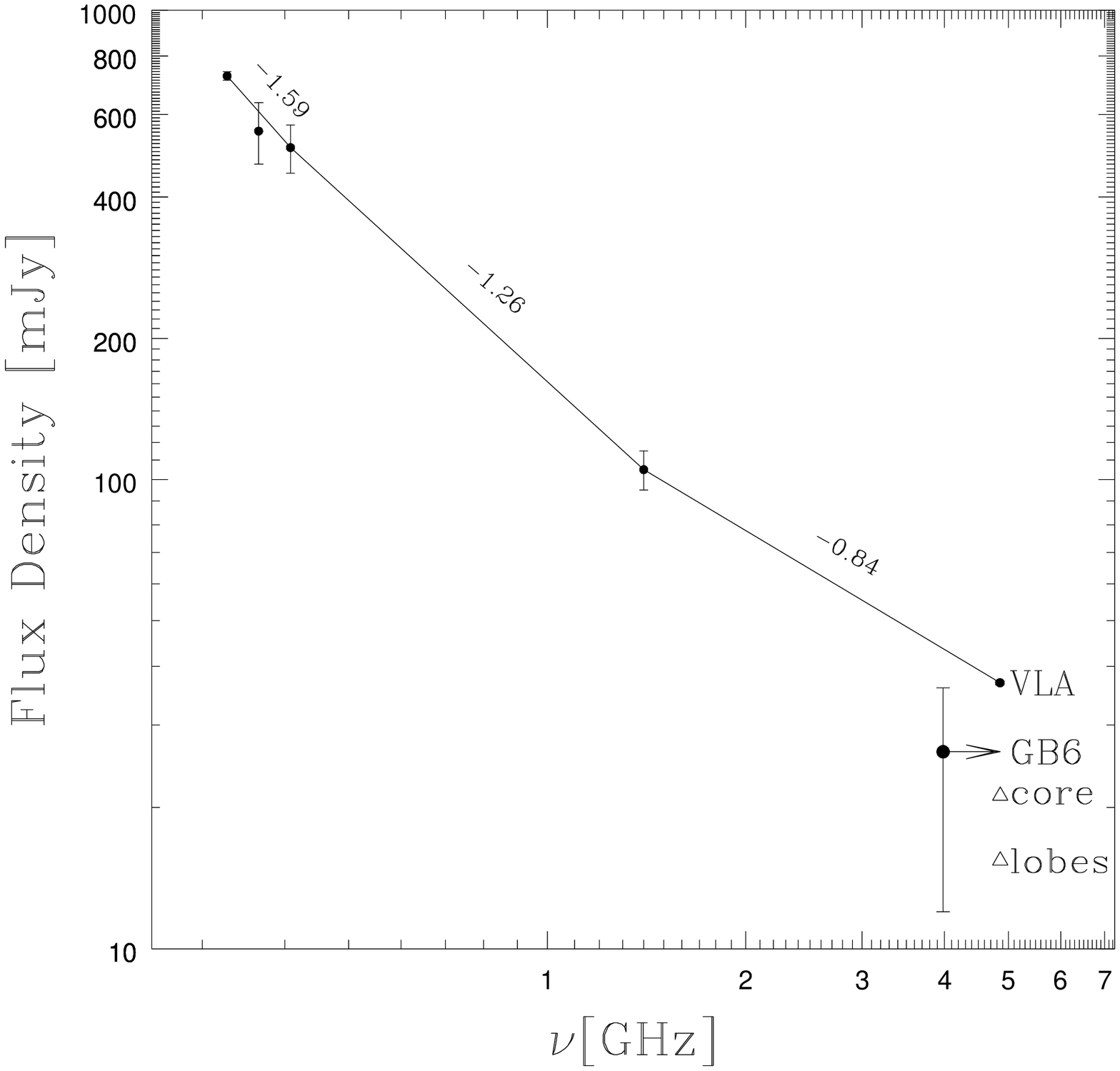,width=16cm}
\figcaption[debreuck.fig2.ps]{Flux density plotted against frequency for the radio data given in Table 1. The dots indicate the total emission, and the triangles the emission from the core and lobes separately. For clarity, the 4.85 GHz GB6 flux has been shifted. Note the flattening of the spectral index with frequency and the excess core emission in our VLA data. \label{debreuck.fig2}}

\psfig{file=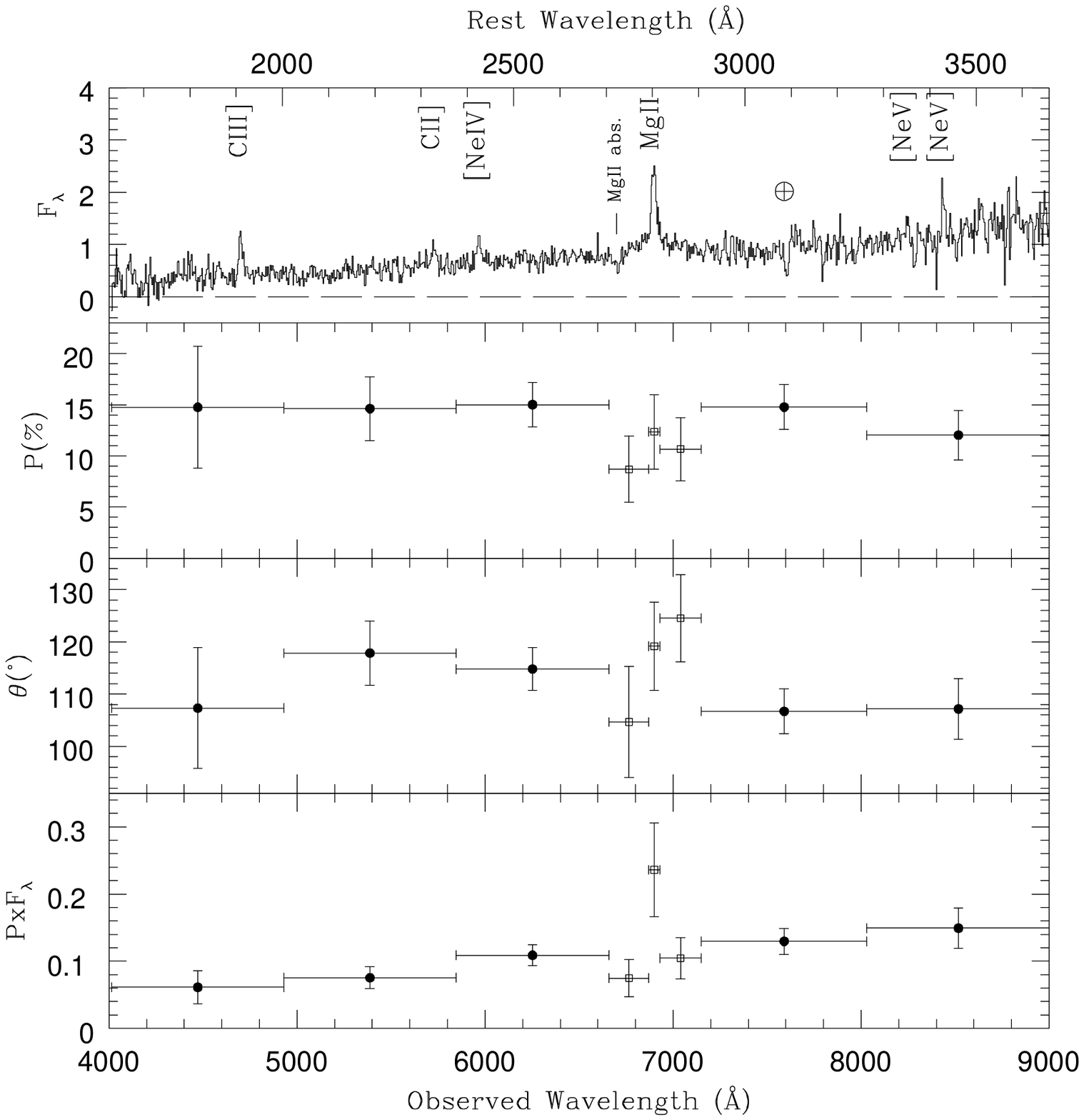,width=14.5cm}
\figcaption[debreuck.fig3.ps]{The wavelength dependence of the total flux density $F_{\lambda}$ units, percentage of polarization $P$, polarization angle $\theta$, and polarized flux $P \times F_{\lambda}$ for WN~J0717+4611 measured in a 6\arcsec $\times$ 1\arcsec~ aperture. $F_{\lambda}$ and $P \times F_{\lambda}$ are in units of 10$^{-18}$ erg s$^{-1}$ cm$^{-2}$ \ang$^{-1}$. $P$, $\theta$ and $P \times F_{\lambda}$ are measured in bins: horizontal bars indicate the bin widths, and vertical bars indicate 1~$\sigma$ errors. The region that has been incompletely corrected for telluric A band absorption is denoted by $\oplus$. Within the errors, the continuum polarization is constant across the spectrum ($P \approx$ 15\%), and similar to that of the emission lines. \label{debreuck.fig3}}

\psfig{file=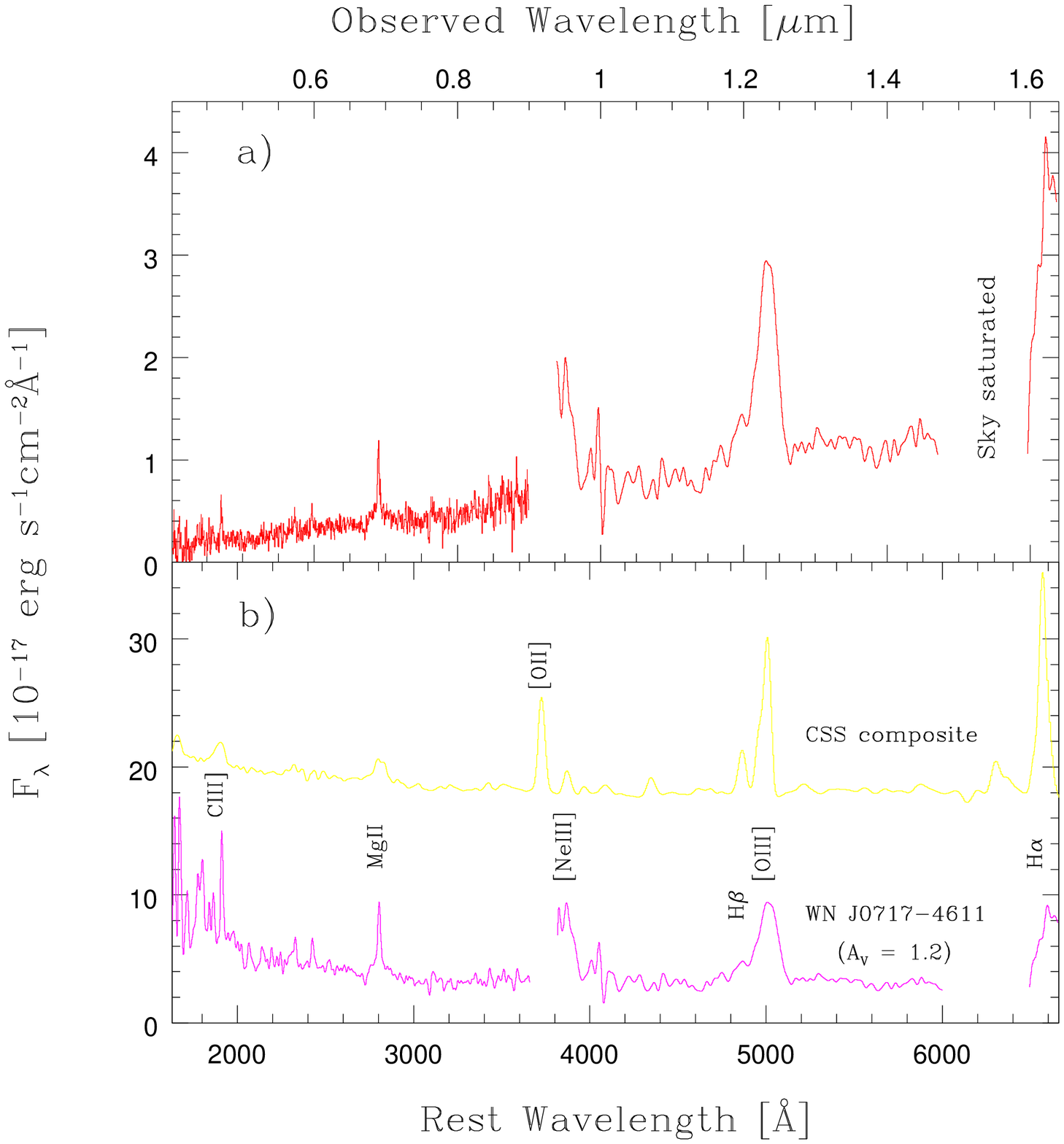,width=15cm}
\figcaption[debreuck.fig4.ps]{a) Optical and IR spectra of WN~J0717+4611, corrected for Galactic reddening with A$_V$=0.24, using the \cite{car89} extinction curve. The flux calibration at both ends of the IR spectrum is highly uncertain, and the region between 14.7 and 16.0 \micron~ is not usable because of saturation by sky lines. b) Top spectrum: Composite CSS spectrum compiled by Baker \& Hunstead (1995), arbitrarily shifted vertically. Bottom spectrum: the spectra in a) dereddened with a local A$_V$ = 1.2 at $z=1.462$ , using an SMC extinction law. Note that the \OII~$\lambda$3727 line falls just between our optical in IR spectra. \label{debreuck.fig4}}

\end{document}